\documentclass[twocolumn,showpacs,aps,superscriptaddress]{revtex4-1}
\usepackage{amsmath}
\usepackage{graphicx}
\usepackage{dcolumn}
\usepackage{float}

\newcommand{\be}{\begin{equation}}
\newcommand{\ee}{\end{equation}}
\newcommand{\bea}{\begin{eqnarray}}
\newcommand{\eea}{\end{eqnarray}}
\newcommand{\bse}{\begin{subequations}}
\newcommand{\ese}{\end{subequations}}
\usepackage{color}

\begin{document}
\title{$\mu$SR and Neutron Diffraction Investigations on Reentrant Ferromagnetic Superconductor Eu(Fe$_{0.86}$Ir$_{0.14}$)$_2$As$_2$ }
\author{V. K. Anand}
\altaffiliation{vivekkranand@gmail.com}
\affiliation{\mbox{Helmholtz-Zentrum Berlin f\"{u}r Materialien und Energie, Hahn-Meitner Platz 1, D-14109 Berlin, Germany}}
\author{D. T. Adroja}
\altaffiliation{devashibhai.adroja@stfc.ac.uk}
\affiliation{ISIS Facility, Rutherford Appleton Laboratory, Chilton, Didcot, Oxon, OX11 0QX, United Kingdom}
\affiliation {\mbox{Highly Correlated Matter Research Group, Physics Department, University of Johannesburg, P.O. Box 524,} Auckland Park 2006, South Africa}
\author{A. Bhattacharyya}
\affiliation{ISIS Facility, Rutherford Appleton Laboratory, Chilton, Didcot, Oxon, OX11 0QX, United Kingdom}
\affiliation {\mbox{Highly Correlated Matter Research Group, Physics Department, University of Johannesburg, P.O. Box 524,} Auckland Park 2006, South Africa}
\author{U. B. Paramanik}
\affiliation {Department of Physics, Indian Institute of Technology, Kanpur 208016, India}
\author{P. Manuel}
\affiliation{ISIS Facility, Rutherford Appleton Laboratory, Chilton, Didcot, Oxon, OX11 0QX, United Kingdom}
\author{A. D. Hillier}
\affiliation{ISIS Facility, Rutherford Appleton Laboratory, Chilton, Didcot, Oxon, OX11 0QX, United Kingdom}
\author{D. Khalyavin}
\affiliation{ISIS Facility, Rutherford Appleton Laboratory, Chilton, Didcot, Oxon, OX11 0QX, United Kingdom}
\author{Z. Hossain}
\altaffiliation{zakir@iitk.ac.in}
\affiliation {Department of Physics, Indian Institute of Technology, Kanpur 208016, India}
\date{\today}

\begin{abstract}
Results of muon spin relaxation ($\mu$SR) and neutron powder diffraction measurements on a reentrant superconductor Eu(Fe$_{0.86}$Ir$_{0.14}$)$_2$As$_2$ are presented. Eu(Fe$_{0.86}$Ir$_{0.14}$)$_2$As$_2$ exhibits superconductivity at $T_{\rm c\,on} \approx 22.5$~K competing with long range ordered Eu$^{+2}$ moments below $\approx 18$~K\@. A reentrant behavior (manifested by nonzero resistivity in the temperature range 10--17.5~K) results from an exquisite competition between the superconductivity and magnetic order. The zero field $\mu$SR data confirm the long range magnetic ordering below $T_{\rm Eu} = 18.7(2)$~K\@. The transition temperature is found to increase with increasing magnetic field in longitudinal field $\mu$SR which along with the neutron diffraction results, suggests the transition to be ferromagnetic.  The neutron diffraction data reveal a clear presence of magnetic Bragg peaks below $T_{\rm Eu}$ which could be indexed with propagation vector {\bf k} = (0,\,0,\,0), confirming a long range magnetic ordering in agreement with $\mu$SR data. Our analysis of the magnetic structure reveals an ordered magnetic moment of $6.29(5)\,\mu_{\rm B}$ (at 1.8~K) on the Eu atoms and they form a ferromagnetic structure with moments aligned along the $c$-axis. No change in the magnetic structure is observed in the reentrant or superconducting phases and the magnetic structure remains same for 1.8~K~$\leq T \leq T_{\rm Eu}$. No clear evidence of structural transition or Fe moment ordering was found.
\end{abstract}

\pacs{74.70.Xa, 75.25.-j, 75.50.cc, 76.75.+i}

\maketitle

\section{\label{Intro} INTRODUCTION}

The discovery of high-temperature superconductivity in 2008 in doped $A{\rm Fe_2As_2}$ ($A$ = Ca, Sr, Ba) after the complete suppression of antiferromagnetic/spin density wave (SDW) transition triggered world-wide research interests in this class of materials \cite{Johnston2010, Canfield2010, Mandrus2010, Stewart2011}. The occurrence of superconductivity was soon found in their magnetic analog, the K-doped EuFe$_2$As$_2$ \cite{Ren2008,Jeevan2008}. EuFe$_2$As$_2$ attracted special attention because of the additional opportunity of exploring the interplay and coexistence of magnetic order and superconductivity brought by the Eu$^{+2}$ ($S = 7/2$) local moments. Like $A{\rm Fe_2As_2}$, EuFe$_2$As$_2$ also crystallizes in the same layered body-centered-tetragonal ${\rm ThCr_2Si_2}$-type structure (space group $I4/mmm$) and exhibits a structural and SDW transition at 190~K associated with the itinerant Fe moments \cite{Johnston2010,Ren2008, Jiang2009a, Xiao2009}. In addition, the Eu$^{+2}$ moments order antiferromagnetically below 19~K with an A-type antiferromagnetic (AFM) structure on Eu sublattice with the ordered Eu$^{+2}$ moments aligned ferromagnetically in $ab$-plane and antiferromagnetically along the $c$-axis \cite{Xiao2009}. The Fe$^{+2}$ moments order antiferromagnetically along the orthorhombic $a$-axis below the structural and SDW transition at 190~K \cite{Xiao2009}. The SDW transition is easily suppressed by partial substitutions at Eu, Fe or As sites or by the application of external pressure leading to superconductivity that coexists with the long range ordering of Eu moments \cite{Jeevan2008, Anupam2011, Jiang2009b, Guguchia2011, Ren2009a, Jeevan2011, Miclea2009, Kurita2011}.

The hole doping by partial K substitution for Eu was found to yield superconductivity in Eu$_{1-x}$K$_{x}$Fe$_{2}$As$_{2}$ with $T_{\rm c} \approx 33$~K for Eu$_{0.5}$K$_{0.5}$Fe$_2$As$_2$ coexisting with short range ordered Eu$^{+2}$ moments below 15~K \cite{Jeevan2008, Anupam2011}. The electron doping by partial Co substitution for Fe leads to reentrant superconductivity in Eu(Fe$_{1-x}$Co$_{x}$)$_{2}$As$_{2}$ with $T_{\rm c} \approx 21$~K and reentrant behavior below 17~K for $x=0.11$ \cite{Jiang2009b,Guguchia2011}. However, no superconductivity is observed in Ni-doped Eu(Fe$_{1-x}$Ni$_{x}$)$_{2}$As$_{2}$, even though the SDW transition is suppressed completely \cite{Ren2009b}. The substitution of Fe by Ni in the optimally hole-doped superconductor Eu$_{0.5}$K$_{0.5}$Fe$_2$As$_2$ was found to lead to reentrant behavior in Eu$_{0.5}$K$_{0.5}$(Fe$_{1-x}$Ni$_{x}$)$_{2}$As$_{2}$ \cite{Anupam2012}. The substitution of As by isovalent P also induces reentrant superconductivity in EuFe$_{2}$(As$_{1-x}$P$_{x}$)$_{2}$ with $T_{\rm c} = 26$~K and Eu magnetic order below 20~K for $x=0.15$ \cite{Ren2009a,Jeevan2011}.  The application of pressure also leads to reentrant superconductivity with $T_{\rm c} \sim 26$~K at a pressure of around 2.5~GPa  \cite{Miclea2009, Kurita2011}.

The interests in these doped EuFe$_2$As$_2$ were sparkled because of the long standing issues of the coexistence of long range magnetic order and superconductivity since such discoveries in rare earth borides and borocarbides \cite{Maple1982,Gupta2006}. In order to understand the interplay of magnetism and superconductivity in these doped EuFe$_2$As$_2$ it is essential to have the knowledge of their magnetic structure. The bulk properties measurements often give an initial idea of possible magnetic structures that needs to be verified by more specific tools such as neutron diffraction (ND) and synchrotron measurements. From the bulk properties measurements of Co-doped EuFe$_2$As$_2$, Jiang et al.\ \cite{Jiang2009b} proposed a helical magnetic structure for Eu(Fe$_{0.89}$Co$_{0.11}$)$_2$As$_2$ and Guguchia et al.\ \cite{Guguchia2011} proposed a canted antiferromagnetic structure for Eu(Fe$_{0.9}$Co$_{0.1}$)$_2$As$_2$. The magnetic structure determination by neutron diffraction study on Eu(Fe$_{0.82}$Co$_{0.18}$)$_2$As$_2$ revealed a ferromagnetic (FM) structure with the ordered Eu$^{+2}$ moments directed along the $c$-axis \cite{Jin2013}. Contradictory antiferromagnetic versus ferromagnetic ground states were reported for P-doped EuFe$_2$As$_2$ by Zapf et al.\ \cite{Zapf2011,Zapf2013} and Nowik et al.\ \cite{Nowik2011} respectively, from the bulk properties measurements and M\"ossbauer studies. Recent x-ray resonant magnetic scattering (XRMS) study by Nandi et al.\ \cite{Nandi2014a} reveal the magnetic structure of reentrant superconductor EuFe$_{2}$(As$_{0.85}$P$_{0.15}$)$_{2}$ to be ferromagnetic with Eu$^{+2}$ moments aligned along the $c$-axis. The ferromagnetic structure of Eu$^{+2}$ moments was further confirmed from the neutron diffraction study on EuFe$_{2}$(As$_{0.81}$P$_{0.19}$)$_{2}$ \cite{Nandi2014b}.

Recently some of us discovered reentrant superconductivity in Ir-doped EuFe$_2$As$_2$ \cite{Paramanik2013,Paramanik2014}. Ir, though isoelectronic to Co, being $5d$ transition metal benefits with an extended $d$ orbital resulting in an increased hybridization and decreased Stoner enhancement factor, therefore the suppression of SDW in parent EuFe$_2$As$_2$ is expected to be more effective with Ir-doping. For the optimal Ir-doping, in Eu(Fe$_{0.86}$Ir$_{0.14}$)$_2$As$_2$, the superconducting transition sets in at $T_{\rm c\,on} \approx 22.5$~K and the long range ordering of Eu$^{+2}$ moments is observed below $\approx 18$~K coexisting with superconductivity \cite{Paramanik2013,Paramanik2014}. The reentrant behavior is reflected by a nonzero resistivity over 10--17.5~K caused by the ordering of Eu$^{+2}$ moments as confirmed by the $^{151}$Eu M\"ossbauer spectroscopy \cite{Paramanik2014}. Investigations on single crystal Eu(Fe$_{0.86}$Ir$_{0.14}$)$_2$As$_2$ revealed evidence for anisotropic magnetic and superconducting properties, the superconducting diamagnetic signal is observed in magnetic susceptibility only for an applied field $H \parallel c$ and not for $H \perp c$, though the $ab$-plane resistivity exhibits zero resistivity state \cite{Paramanik2014}.  Jiao et al.\ suggested the nature of magnetic ordering to be ferromagnetic in Eu(Fe$_{0.88}$Ir$_{0.12}$)$_2$As$_2$ \cite{Jiao2013}. Isoelectronic substitution of Fe by Ru is also reported to have ferromagnetic ordering of Eu-moment in Eu(Fe$_{1-x}$Ru$_{x}$)$_{2}$As$_{2}$ coexisting with superconductivity \cite{Jiao2011}. The nonsuperconducting Eu(Fe$_{1-x}$Ni$_{x}$)$_{2}$As$_{2}$ also have ferromagnetic ground state for $0.03\leq x \leq 0.1$ \cite{Ren2009b}.

The neutron diffraction study on Eu(Fe$_{0.82}$Co$_{0.18}$)$_2$As$_2$ \cite{Jin2013} and EuFe$_{2}$(As$_{0.81}$P$_{0.19}$)$_{2}$ \cite{Nandi2014b} provide clear evidence for the coexistence of ferromagnetism and superconductivity in these systems.  While the coexistence of antiferromagnetic order and superconductivity is now perceived, the coexistence of ferromagnetic order and superconductivity which are antagonistic to each other is still puzzling and deserves microscopic investigations. In order to check the above inference of ferromagnetic ordering in Ir-doped EuFe$_2$As$_2$ and find out whether the ferromagnetic nature of Eu-magnetic order is more general for the doped EuFe$_2$As$_2$ or specific to the case of substitution of Fe by Co and of As by P we have determined the magnetic structure of reentrant superconductor Eu(Fe$_{0.86}$Ir$_{0.14}$)$_2$As$_2$. Here we report our muon spin relaxation ($\mu$SR) and neutron powder diffraction  investigations on Eu(Fe$_{0.86}$Ir$_{0.14}$)$_2$As$_2$ and show that the nature of the long range magnetic order of Eu$^{+2}$ moments is indeed ferromagnetic for Ir-doping too.

Theoretically ferromagnetism has been suggested to coexist with superconductivity under certain conditions: (a) when the ferromagnetism is cryptoferromagnetic with multidomains such that the net magnetization is zero over the superconducting coherence length \cite{Anderson1959}, (b) when ferromagnetism polarizes the Cooper pairs leading to an inhomogeneous superconductivity referred as Fulde-Ferrell-Larkin-Ovchinikov (FFLO) phase \cite{Fulde1964,Larkin1965}, and (c) when a spontaneous-vortex phase is formed by the internal field of ferromagnetic order such that the combined energy of the coexisting phase is lowered \cite{Greenside1981}. Jiao et al.\ suggested the possibility of spontaneous-vortex phase and/or FFLO state for the coexistence of ferromagnetic order and superconductivity in Eu(Fe$_{0.75}$Ru$_{0.25}$)$_2$As$_2$ \cite{Jiao2011}. Nandi et al.\ suggested spontaneous-vortex in EuFe$_{2}$(As$_{0.85}$P$_{0.15}$)$_{2}$ \cite{Nandi2014a}.  The coexistence of ferromagnetism and superconductivity in Eu(Fe$_{0.86}$Ir$_{0.14}$)$_2$As$_2$ may also have similar mechanism.

In the following we present our results of $\mu$SR and ND measurements on Eu(Fe$_{0.86}$Ir$_{0.14}$)$_2$As$_2$. The zero field (ZF) $\mu$SR results confirm the long range magnetic ordering below $T_{\rm Eu} = 18.7(2)$~K and the longitudinal field (LF) $\mu$SR indicated the nature of long range order to be ferromagnetic. The neutron diffraction data allowed us to determine the magnetic structure, further supporting the ferromagnetic order below $T_{\rm Eu}$. The ordered Eu$^{+2}$ moments lie along the tetragonal $c$-axis with a magnetic propagation wavevector {\bf k} = (0,\,0,\,0). At 1.8~K the ordered state moment is found to be $6.29(5)\,\mu_{\rm B}$. The $\mu$SR or neutron diffraction data do not show any evidence for any change in magnetic structure in superconducting and reentrant phases.

\section{\label{ExpDetails} EXPERIMENTAL DETAILS}

The polycrystalline sample of Eu(Fe$_{0.86}$Ir$_{0.14}$)$_2$As$_2$ was prepared by the solid state reaction method \cite{Paramanik2013,Paramanik2014} starting with the high purity elements (Eu: 99.9\%, Fe: 99.999\%, Ir: 99.99\%, and As: 99.999\%) in stoichiometric ratio. Fine pieces of Eu were mixed with Fe, Ir and As powders, pelletized and sealed in an evacuated quartz tube and fired at 900~$^\circ$C for 30~h\@. After the first firing the sample was thoroughly ground and again pelletized and sealed in evacuated quartz tube. The sintering for 5~days at 900~$^\circ$C yielded good quality sample as indicated by the powder x-ray diffraction (XRD) data.

The $\mu$SR measurement was carried out at the ISIS facility of the Rutherford Appleton Laboratory, Didcot, U.K.\ using the EMU spectrometer both in zero field and in longitudinal fields up to 0.45~T\@. The powdered sample of Eu(Fe$_{0.86}$Ir$_{0.14}$)$_2$As$_2$ was mounted on a high purity silver plate using diluted GE varnish and covered with kapton film. The low temperature was achieved by cooling the sample in a standard He-4 cryostat. In order to minimize the effect of the stray fields at the sample position the correction coils were used which ensured the stray field to be within 1~$\mu$T. The zero-field $\mu$SR data were collected in the temperature range 1.2~K to 100~K, and LF data were collected at 1.4~K, 10~K, 18~K and 35~K for fields up to 0.45~T\@.

The neutron diffraction measurements were carried out using the WISH time of flight diffractometer \cite{Chapon2011} at the ISIS Facility. The powdered Eu(Fe$_{0.86}$Ir$_{0.14}$)$_2$As$_2$ sample was lightly packed in a thin-walled vanadium can (diameter 3~mm) to reduce packing ratio and thereby absorption. The low temperature was achieved by cooling the sample inside a He-4 cryostat using He-exchange gas to ensure good thermal contact at low temperature. ND data were collected with long counting (4 hours per run) at 1.8~K, 12~K, 16~K and 25~K in order to determine magnetic structure as well as any change in magnetic structure with temperature. We also collected data at several temperatures between 1.8~K and 25~K with shorter counting time (30 min per point) to investigate the temperature dependence of order parameter. The refinement of neutron diffraction data was carried out using Fullprof program \cite{Rodriguez1993}.

\section{\label{Sec:muSR} Muon Spin Relaxation Study}

\begin{figure}
\includegraphics[width=3in]{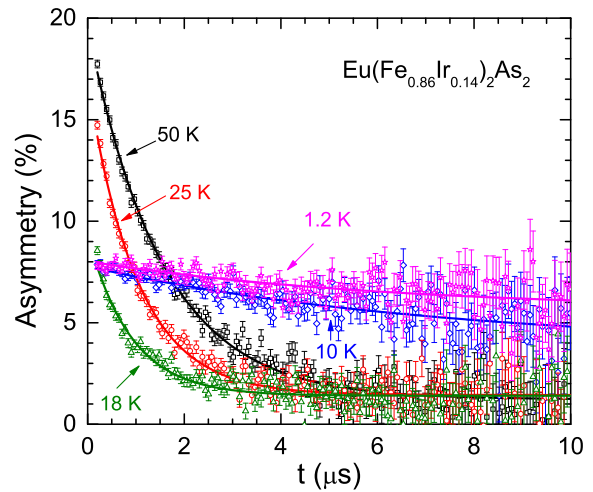}
\caption{\label{fig:MuSR1} (Color online) Zero field muon spin asymmetry function $G_z(t, H)$ versus time $t$ $\mu$SR spectra of Eu(Fe$_{0.86}$Ir$_{0.14}$)$_2$As$_2$ at few representative temperatures. Solid curves are the fits to the $\mu$SR data by the relaxation function in Eq.~(\ref{eq:muSR}). }
\end{figure}

\begin{figure}
\includegraphics[width=3in]{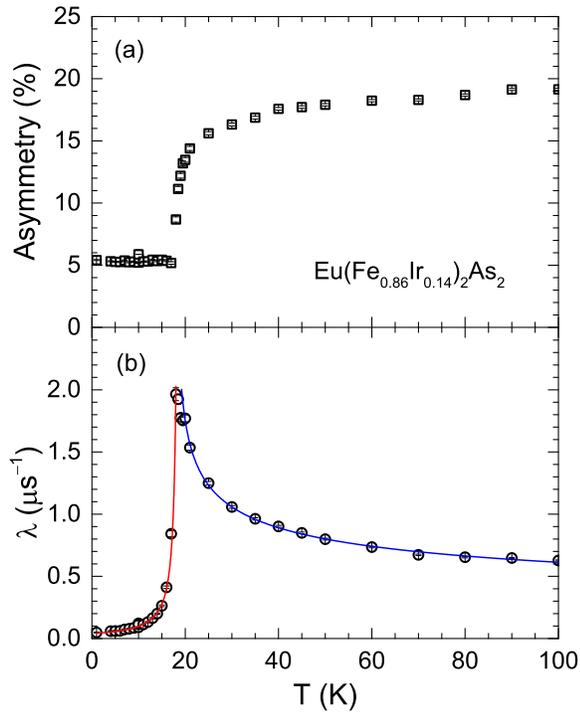}
\caption{\label{fig:MuSR2} (Color online) Temperature $T$ dependence of (a) the initial asymmetry $A_0$ and (b) the depolarization rate $\lambda$ obtained from the analysis of the zero field $\mu$SR data of Eu(Fe$_{0.86}$Ir$_{0.14}$)$_2$As$_2$ for $1.2~{\rm K} \leq T \leq 100$~K\@. The solid curves in (b) are the fits to critical exponent behavior $\lambda(T) = \lambda_0 |1 - T/T_{\rm Eu}|^{-w}$. }
\end{figure}

In order to shed light on the magnetic phase transitions seen in the zero-field resistivity of Eu(Fe$_{0.86}$Ir$_{0.14}$)$_2$As$_2$ below the superconducting transition $T_{\rm c} \approx 21.5$~K, we have investigated the temperature $T$ dependence of the muon spin relaxation in zero-field (ZF), in applied longitudinal magnetic fields $H= 0.10$~T and 0.45~T as well as $H$ dependence at $T=10$~K and 18~K\@. Figure~\ref{fig:MuSR1} shows the $\mu$SR asymmetry spectra of Eu(Fe$_{0.86}$Ir$_{0.14}$)$_2$As$_2$ at few representative temperatures between 1.2~K and 50~K collected in zero field while warming the sample. For $T \geq 20~{\rm K}$, the $\mu$SR spectra exhibit a typical behaviour expected from the fluctuating paramagnetic moments with full initial asymmetry of 20\% at 100~K\@. At $T<20$~K a loss ($\sim 2/3$ at 1.2~K) in the initial asymmetry is clearly seen which is an indication of a long range magnetic ordering.  Further the absence of asymmetry-time oscillations indicates that the ordered state moments are too large to be observed in the time windows of the EMU spectrometer.

The ZF $\mu$SR spectra were fitted using a Lorentzian (also called exponential function) decay,
\begin{equation}
 G_z(t,H) = A_0 \exp(-\lambda t) + A_{\rm BG}
 \label{eq:muSR}
\end{equation}
where $A_0$ is the initial asymmetry parameter, $\lambda$ is the electronic relaxation rate mainly arising from the local  moments and $A_{\rm BG}$ is a nonrelaxing constant background from the silver sample holder. The $A_{\rm BG}$  was estimated from the 100~K ZF data and was kept fixed for the rest of the analysis. The fits of the $\mu$SR data by the relaxation function in Eq.~(\ref{eq:muSR}) are shown by the solid curves in Fig.~\ref{fig:MuSR1}. Figure~\ref{fig:MuSR2} shows the temperature dependence of parameters $A_0$ and $\lambda$ obtained from the fits of the ZF $\mu$SR spectra for $1.2~{\rm K} \leq T \leq 100$~K. It is seen from Fig.~\ref{fig:MuSR2}(a) that the initial asymmetry $A_0$ starts decreasing as $T$ is lowered below 24~K and exhibits a sharp drop below 20~K\@. At low-$T$ $A_0$ is only 1/3 of its high temperature value, i.e. there is a drop in $A_0$  by 2/3 of its high-$T$ value, which confirms the bulk nature of the long range magnetic ordering in Eu(Fe$_{0.86}$Ir$_{0.14}$)$_2$As$_2$. Further, $\lambda$ increases with decreasing temperature below 100~K and exhibits a peak near 19~K, which is due to the long range magnetic ordering. The absence of any additional anomaly both in $A_0$ and $\lambda$ below 19~K, indicates that there is only one magnetic phase transition at $T_{\rm Eu} \approx 19$~K which is probed by muon spin relaxation. Thus we do not see reentrant feature in $\mu$SR.

It is interesting to note from Fig.~\ref{fig:MuSR2}(a) that the magnetic volume fraction of sample is close to 100\%. The superconducting volume fraction has also been found to be nearly 100\% from the magnetic susceptibility measurements on Eu(Fe$_{0.86}$Ir$_{0.14}$)$_2$As$_2$ \cite{Paramanik2013,Paramanik2014}. This suggests that the whole volume of the sample participates in both magnetic ordering and superconductivity.

\begin{figure}
\includegraphics[width=\columnwidth]{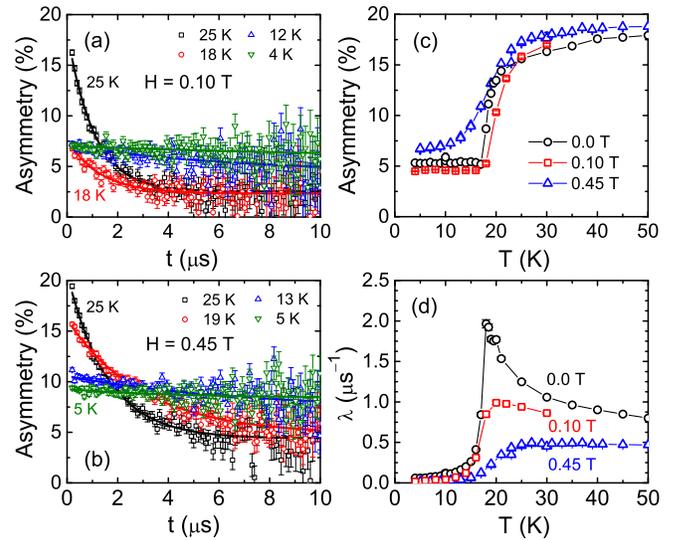}
\caption{\label{fig:MuSR3} (Color online) Longitudinal field (LF) muon spin asymmetry function $G_z(t, H)$ versus time $t$ $\mu$SR spectra of Eu(Fe$_{0.86}$Ir$_{0.14}$)$_2$As$_2$ at few representative temperatures for fields (a) $H= 0.10$~T and (b) $H= 0.45$~T\@. Solid curves are the fits to the $\mu$SR data by the relaxation function in Eq.~(\ref{eq:muSR}). The temperature $T$ dependence of (c) the initial asymmetry $A_0$ and (d) the depolarization rate $\lambda$ obtained from the analysis of the LF $\mu$SR data at $H= 0.10$~T and 0.45~T for $4~{\rm K} \leq T \leq 45$~K\@. The zero field $A_0(T)$ and $\lambda(T)$ are also shown for comparison.}
\end{figure}

\begin{figure}
\includegraphics[width=\columnwidth]{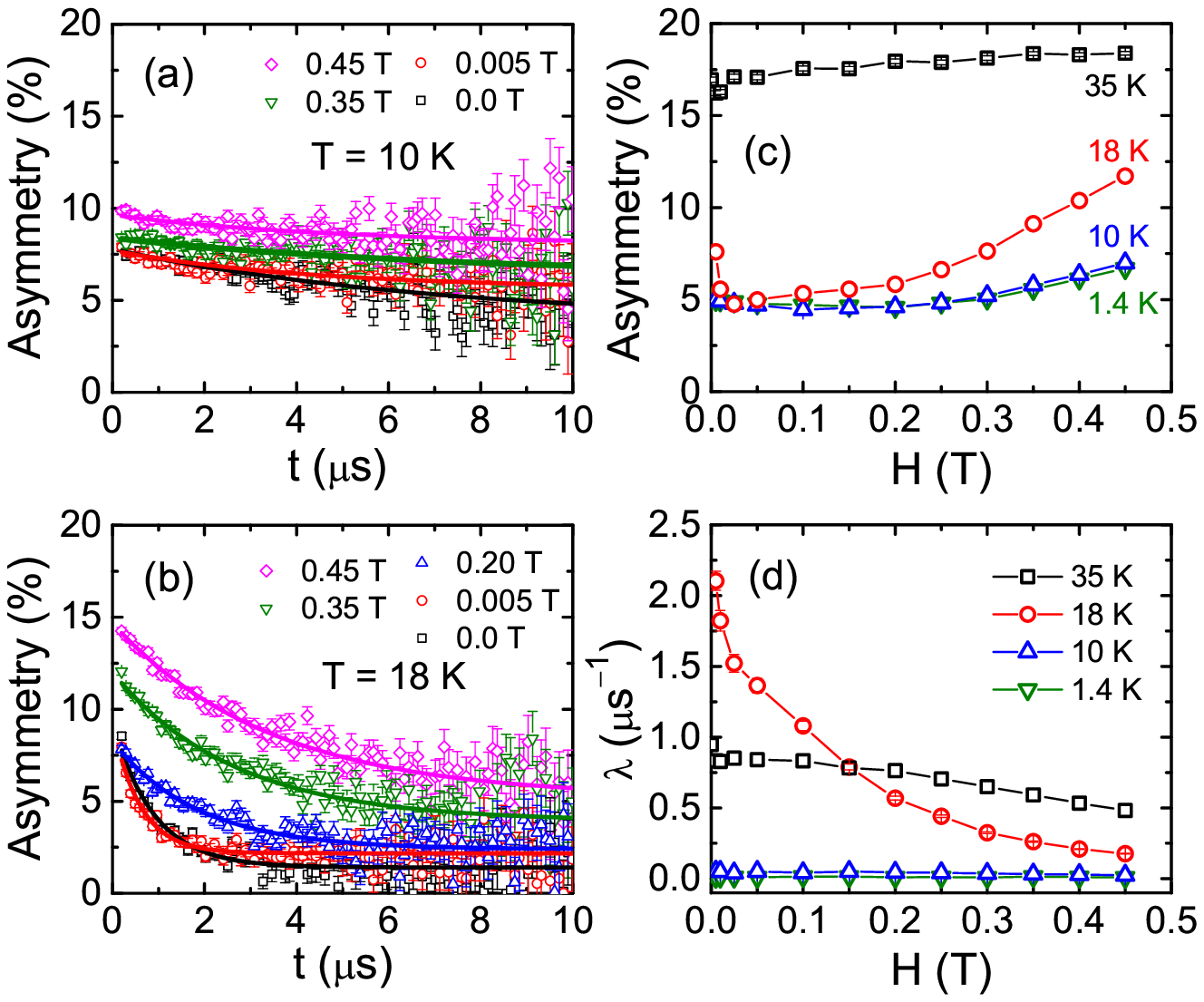}
\caption{\label{fig:MuSR4} (Color online) Longitudinal field (LF) muon spin asymmetry function $G_z(t, H)$ versus time $t$ $\mu$SR spectra of Eu(Fe$_{0.86}$Ir$_{0.14}$)$_2$As$_2$ at temperatures (a) $T= 10$~K and (b) $T= 18$~K for few representative fields. Solid curves are the fits to the $\mu$SR data by the relaxation function in Eq.~(\ref{eq:muSR}). The magnetic field $H$ dependence of (c) the initial asymmetry $A_0$ and (d) the depolarization rate $\lambda$ obtained from the analysis of the LF $\mu$SR data for $0 \leq H \leq 0.45$~T\@.}
\end{figure}

Though, $\lambda$ is not a strict order parameter, it is a measure of spatial correlation length $\xi$ and can give an idea of critical dynamics: $\lambda\sim \xi^{3/2}$ for a ferromagnet and $\lambda\sim \xi^{1/2}$ for an antiferromagnet, whereas $\xi \sim (T/T_{\rm tr}-1)^{-\nu}$ in the close vicinity of the transition temperature $T_{\rm tr}$  \cite{Lovesey1995a, Lovesey1995b,Yaouanc1993}.  The $T$ dependence of $\lambda$ could be described well by the critical exponent behavior $\lambda(T) = \lambda_0 |1 - T/T_{\rm Eu}|^{-w}$  \cite{Yaouanc1993,Cywinski1991,Henneberger1999} for both $T< T_{\rm Eu}$ and $T> T_{\rm Eu}$. For a 3-dimensional (3D) ferromagnet (neutron diffraction data in next section reveal a 3D ferromagnetism in the present compound) the exponent $w \approx \nu (z-1) \approx 1.05$ with exponents $\nu \approx 0.70$ and $z=5/2$  \cite{Yaouanc1993,Henneberger1999,Hohenemser1989}. The fits of $\lambda(T)$ are shown by the solid curves in Fig.~\ref{fig:MuSR2}(b). For $T< T_{\rm Eu}$ a good fit is obtained for fit parameters $\lambda_0 = 0.039(2)~\mu$s$^{-1}$, $T_{\rm Eu} = 18.7(2)$~K and $w = 1.20(6)$ from fitting over $1.2~{\rm K} \leq T \leq 18$~K\@. Whereas for $T> T_{\rm Eu}$ the fit parameters are $\lambda_0 = 0.94(2)~\mu$s$^{-1}$, $T_{\rm Eu} = 18.0(5)$~K and $w = 0.28(2)$ for fit over $20~{\rm K} \leq T \leq 100$~K\@. It is seen that for $T< T_{\rm Eu}$  we obtain exponent $w = 1.20(6)$ which is close to the expected value of  $w\approx 1.05$  for 3D ferromagnet. A similar value of the exponent $w = 1.06(9)$ is reported for the ferrimagnet Cu$_2$OSeO$_3$ which was obtained from the fit of $\lambda(T)$ in the paramagnetic state ($ T> T_{\rm c}$)  \cite{Maisuradze2011}. However, for $T> T_{\rm Eu}$ we see that the exponent $w$ is much lower than the expected value. Even for the fitting range close to $T_{\rm Eu}$ in $T \leq 1.3 T_{\rm Eu}$ there is no gain in $w$. This deviation possibly could be due to the muon-lattice dipolar interaction which is known to strongly affect the paramagnetic critical dynamics near the transition temperature of a ferromagnet \cite{Yaouanc1993,Henneberger1999}. Efforts have been made in past to determine the $T$ dependence of $\lambda$ in the critical regime of dipolar Heisenberg ferromagnets using the mode coupling (MC) theory \cite{Lovesey1995a, Yaouanc1993, Frey1994,Yaouanc1996}. Within the MC theory the effect of dipolar interactions is accounted for by the splitting of magnetic fluctuations into longitudinal and transverse modes in the reciprocal space, and the relative weight of transverse and longitudinal modes determines the critical $\lambda(T)$ behavior. Unfortunately, we do not have all the necessary ingredients for the calculation of MC theory prediction for our compound. We compare our results with the MC theory prediction for the 3D Heisenberg ferromagnet EuO \cite{Yaouanc1993,Blundell2010}. From Fig.~\ref{fig:MuSR2}(b) we see that there is a noticeable rise in $\lambda$ as the temperature is lowered from 50~K towards $T_{\rm Eu}$ with a maximum value of $\lambda \approx 2~\mu$s$^{-1}$ near $T_{\rm Eu}$. This value is very small compared to MC theory calculated critical value of $\sim 12~\mu$s$^{-1}$ \cite{Yaouanc1993}, but very close to the experimental value of $\lambda \approx 2~\mu$s$^{-1}$ for EuO near $T_{\rm c}=70$~K \cite{Blundell2010}.

The LF $\mu$SR measurements in $H= 0.10$~T and 0.45~T were done to understand the nature of the magnetic phase transition at $T_{\rm Eu} \approx 19$~K\@. The $H= 0.10$~T and 0.45~T LF $\mu$SR data at various $T$ were also fitted by the relaxation function in Eq.~(\ref{eq:muSR}), thus yielding the $T$ and $H$ dependent fit parameters. The LF $\mu$SR data at selected temperatures and  their fits are shown in Figs.~\ref{fig:MuSR3}(a) and (b). The $T$ dependence of the fit parameters $A_0$ and $\lambda$ obtained so for $H= 0.10$~T and 0.45~T are shown in Figs.~\ref{fig:MuSR3}(c) and (d) and compared with the ZF $A_0(T)$ and $\lambda(T)$. It is seen from Fig.~\ref{fig:MuSR3}(c) that while at $H = 0.10$~T the LF $A_0(T)$ is nearly similar to ZF $A_0(T)$, at $H = 0.45$~T there is a small increase in $A_0$. A substantial influence of field is observed on the $T$ dependence of $\lambda$ [Fig.~\ref{fig:MuSR3}(d)]. At $H=0.10$~T the peak height of $\lambda$ is decreased and at $H=0.45$~T the peak in $\lambda(T)$ almost disappears. Further at $H=0.45$~T the drop in the $\lambda(T)$ starts at a temperature around 25~K compared to 19~K for ZF. This indicates that at $H=0.45$~T the transition temperature has increased to 25~K from 19~K, thus revealing a ferromagnetic nature of the phase transition at $T_{\rm Eu} \approx 19$~K as deduced from the neutron diffraction data in the next section.

In Figs.~\ref{fig:MuSR4}(a) and (b) we show the LF $\mu$SR asymmetry spectra of Eu(Fe$_{0.86}$Ir$_{0.14}$)$_2$As$_2$ at different fields at $T= 10$~K and 18~K\@. It is seen that the initial asymmetry increases with increasing field. While the increase in initial asymmetry is very weak at 10~K, the increase is significant at 18~K, increases by almost a factor of 2 at 0.45~T\@. The LF $\mu$SR data at fields $0\leq H \leq 0.45$~T at $T=1.4$~K, 10~K, 18~K and 35~K were fitted by the relaxation function in Eq.~(\ref{eq:muSR}). The representative fits are shown in Figs.~\ref{fig:MuSR4}(a) and (b). The fit parameters $A_0$ and $\lambda$ are shown in Figs.~\ref{fig:MuSR4}(c) and (d) as a function of $H$\@. It is seen from Fig.~\ref{fig:MuSR4}(c) that at 1.4~K and 10~K $A_0$ shows a weak increase with increasing $H$ for $H\geq 0.2$~T\@. This increase in $A_0$ for $H\geq 0.2$~T is quite significant at 18~K. At 35~K too an extremely weak increase in $A_0$ is observed. On the other hand the $\lambda$ is almost insensitive to field at 1.4~K and 10~K [Fig.~\ref{fig:MuSR4}(d)]. At 18~K  $\lambda$ shows strong dependence on $H$.  The $\lambda$ decreases with increasing $H$ for the entire range of $H$ ($0 \leq H \leq 0.45$~T)\@. A weak decrease in $\lambda$ at $H\geq 0.2$~T is observed even at 35~K\@.

\section{\label{Sec:neutrom} Neutron Diffraction Study}
\begin{figure}
\includegraphics[width=3in]{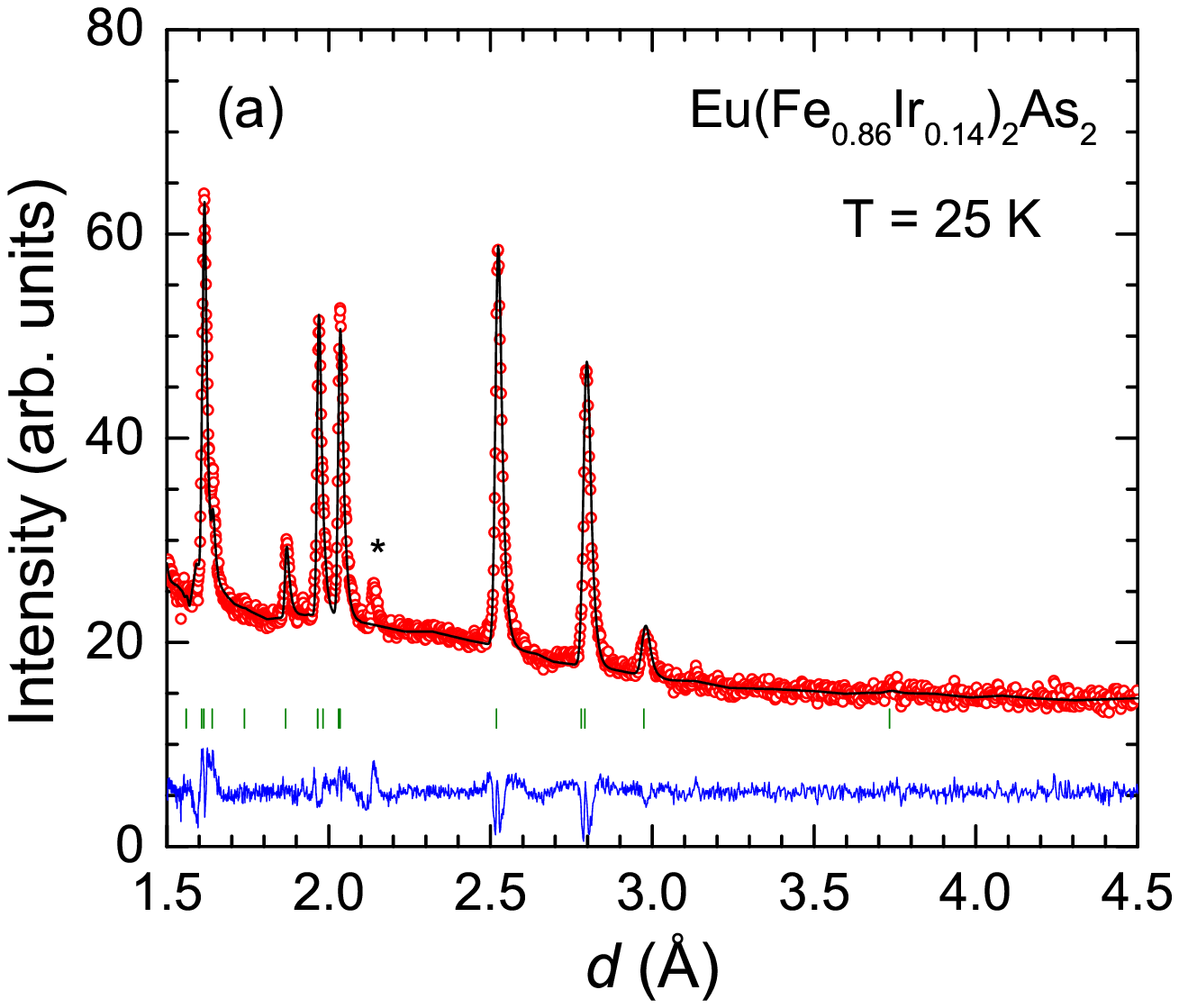}
\includegraphics[width=3in]{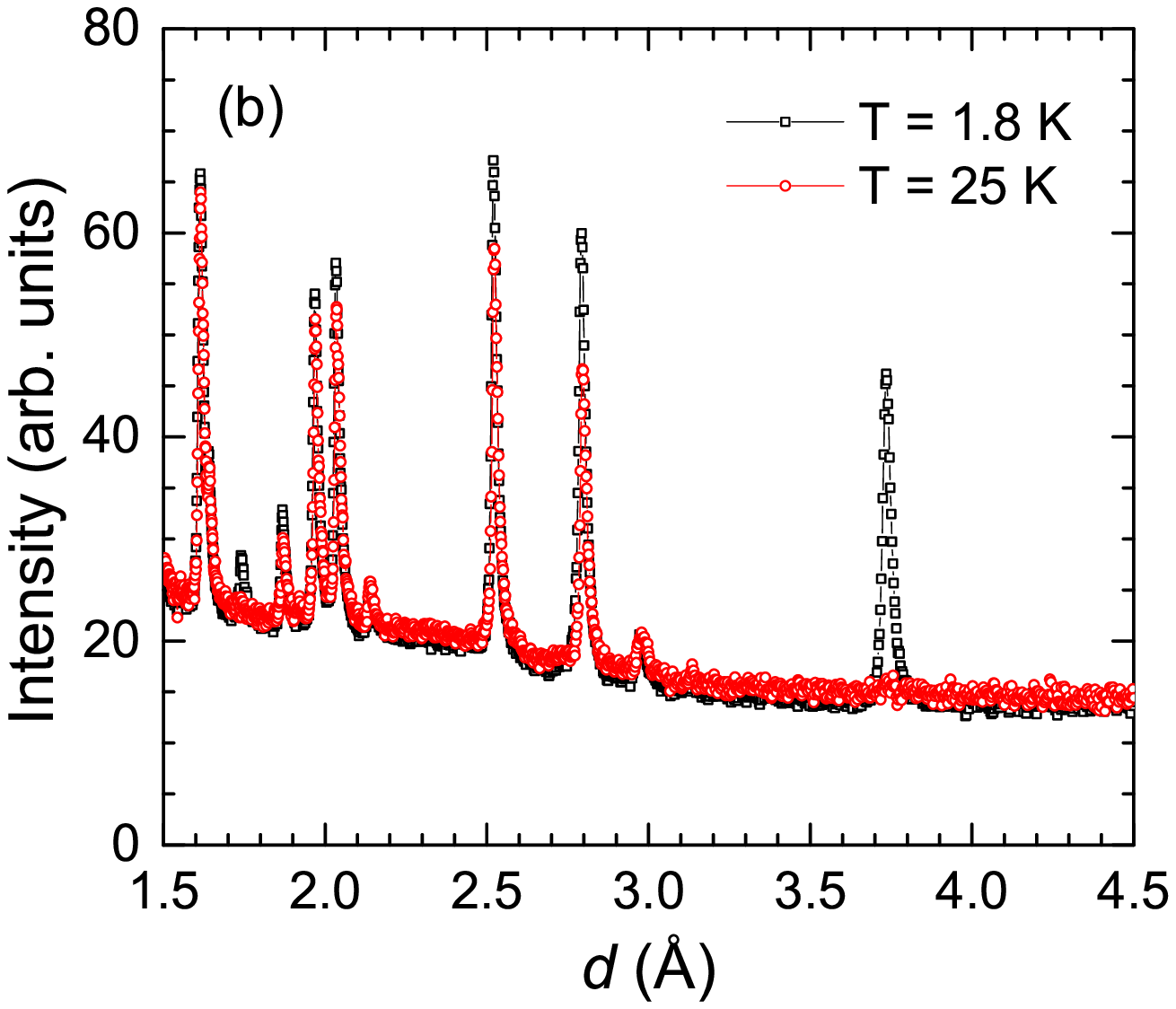}
\includegraphics[width=3in]{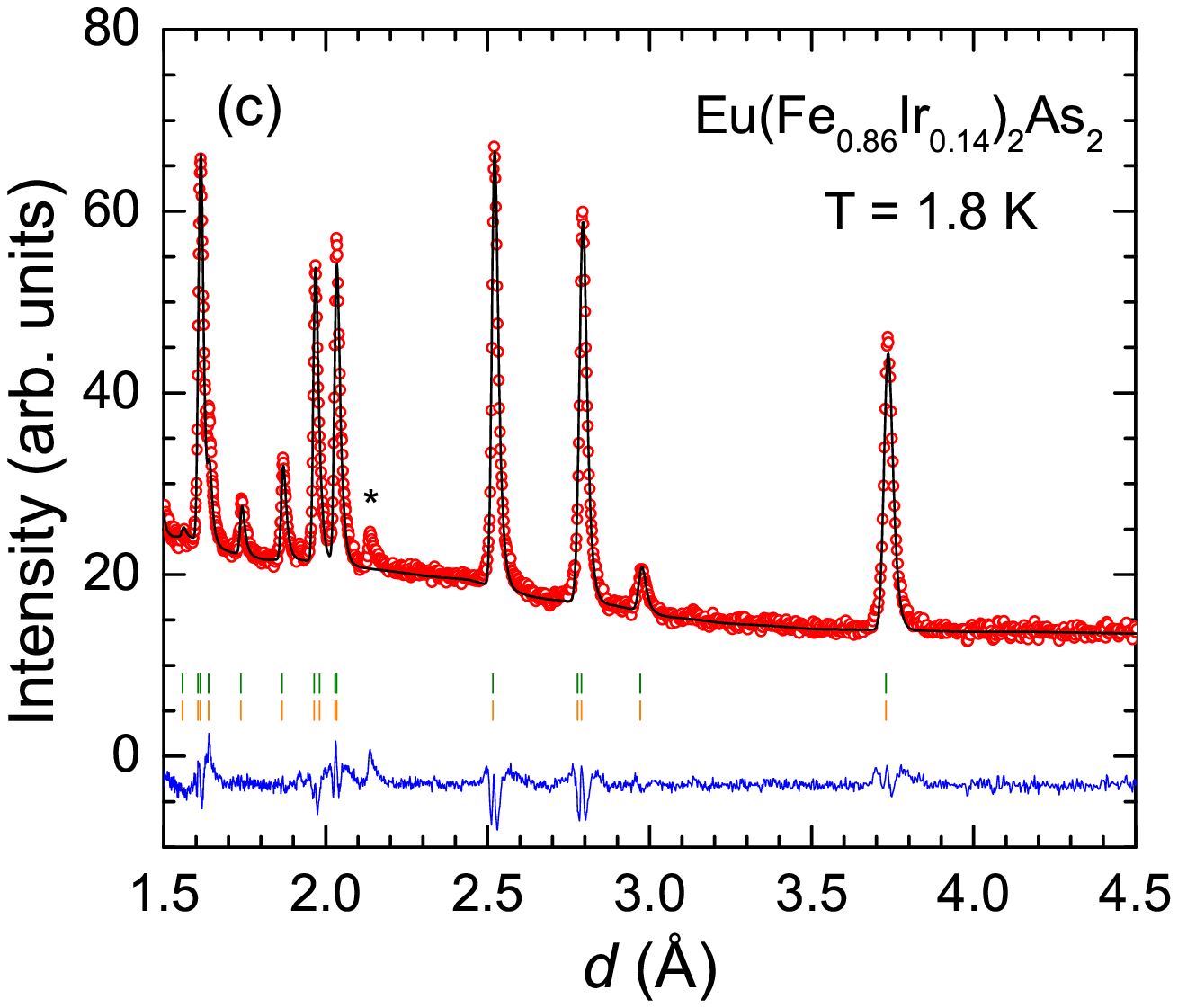}
\caption{\label{fig:ND1} (Color online) (a) Neutron diffraction (ND) pattern of Eu(Fe$_{0.86}$Ir$_{0.14}$)$_2$As$_2$ recorded at 25 K. The solid line through the experimental points
is the structural Rietveld refinement profile calculated for the ThCr$_2$Si$_2$-type body centered tetragonal (space group $I4/mmm$) structure. The vertical bars indicate the Bragg peak positions. The lowermost curve represents the difference between the experimental data and calculated intensities. (b) Comparison of ND patterns recorded at 1.8~K and 25 K\@. (c) The structural and magnetic refinement profile for ND pattern at 1.8~K\@. Peaks marked with star in (a) and (c) arise from the sample holder.}
\end{figure}

Figure~\ref{fig:ND1} shows the neutron diffraction (ND) data collected at 1.8 K and 25 K for the detector bank-3 (centered at 90$^\circ$) of WISH. The Fullprof structural refinement of ND data at 25~K reveal ThCr$_2$Si$_2$-type body centered tetragonal (space group $I4/mmm$) structure of Eu(Fe$_{0.86}$Ir$_{0.14}$)$_2$As$_2$. The refinement profile is shown in Fig.~\ref{fig:ND1}(a) and crystallographic parameters are listed in Table~\ref{tab:ND1}. Since we did not observe any clear impurity phase we expect the chemical composition of the sample to be the same as the starting stoichiometry. Accordingly, while refining, the occupancies of Fe and Ir were kept fixed according to the stoichiometric ratio. The crystallographic parameters obtained from refinement of 25~K ND data are compared with those obtained from the refinement of room temperature powder XRD data in Table~\ref{tab:ND1}. While the lattice parameter $a$ and As $c$ axis coordinate $z_{\rm As}$ are nearly same, the lattice parameter $c$ and hence unit cell volume $V_{\rm cell}$ appears to decrease at lower temperature. The ND data thus show that the crystal structure of Eu(Fe$_{0.86}$Ir$_{0.14}$)$_2$As$_2$ at 25~K remains the same as at room temperature. Furthermore, no evidence of structural change is obtained from the ND data down to 1.8~K\@.

\begin{table}
\caption{Crystallographic and Rietveld refinement parameters obtained from room temperature powder XRD and 25~K neutron diffraction data of Eu(Fe$_{0.86}$Ir$_{0.14}$)$_2$As$_2$ with the body-centered tetragonal ${\rm ThCr_2Si_2}$-type structure (space group  $I4/mmm$). Also listed are the parameters obtained from the crystal and magnetic structures refinement of 1.8~K neutron diffraction data. The atomic coordinates of Eu, Fe/Ir and As atoms are (0,0,0), (0,1/2,1/4) and (0,0,$z_{\rm As}$), respectively.}
\label{tab:ND1}
\begin{ruledtabular}
\begin{tabular}{lccc}
 & XRD  & ND  & ND \\
  & (300~K) & (25~K) & (1.8~K)\\
\hline
\underline{Lattice parameters}\\
\hspace{0.8cm} $a$ (\AA)            		&  3.9365(8) &3.9326(4) & 3.9287(3) \\	
\hspace{0.8cm} $c$ (\AA)          			&  12.027(4) &11.897(1) & 11.884(2)\\
\hspace{0.8cm} $V_{\rm cell}$  (\AA$^3$) 	&  186.37(7) & 183.99(2)& 183.42(3) \\
\underline{Atomic coordinate}\\
 \hspace{0.8cm} $z_{\rm As}$   &  0.3613(6) & 0.3606(4) & 0.3604(4)\\
\underline{Refinement quality} \\
\hspace{0.8cm} $\chi^2$          & 1.33 & 1.98 & 2.75\\	
\hspace{0.8cm} $R_{\rm p}$ (\%)  & 15.8 & 7.39 & 4.13\\
\hspace{0.8cm} $R_{\rm wp}$ (\%) & 20.5 & 4.96 & 4.28\\
\end{tabular}
\end{ruledtabular}
\end{table}

A comparison of ND data at 25~K and 1.8~K is shown in Fig.~\ref{fig:ND1}(b). It is clearly seen that at low temperature (1.8~K) there is a significant increase in the intensity of several nuclear peaks, in particular there is a very strong enhancement in the intensity of (1\,0\,1) nuclear peak at $d=3.73$~\AA\ (which is almost zero at 25~K), which as expected, indicates a long range  magnetic phase transition. The $Q$-dependence ($Q=2\pi/d$) of the intensities of these peaks (strong at smaller-$Q$ or at larger $d$-spacing) confirms that they are due to the long range magnetic ordering of the Eu/Fe-moment. Further, it is seen that all the magnetic peaks appear only at the position of nuclear peaks which can be taken as a signature of ferromagnetic nature of ordering. No additional magnetic Bragg peaks were observed in the reentrant state (e.g. at 14~K, data not shown). Therefore the magnetic structure seems to remain unchanged between the reentrant and superconducting phases.

In order to estimate the magnetic propagation vector {\bf k}, an automatic indexing procedure using a grid search in Fullprof program was used. Our neutron diffraction data provide a direct estimation of the propagation vector, which is {\bf k} = (0,\,0,\,0).  The {\bf k} = (0,\,0,\,0) propagation vector has also been seen for ferromagnetic ordering of the Eu$^{+2}$ moments in Eu(Fe$_{0.82}$Co$_{0.18}$)$_2$As$_2$ \cite{Jin2013} and EuFe$_{2}$(As$_{0.81}$P$_{0.19}$)$_{2}$ \cite{Nandi2014b}. For the parent compound EuFe$_2$As$_2$ the A-type antiferromagnetism of Eu$^{+2}$ is characterized by the propagation vector {\bf k} = (0,\,0,\,1) and the ordering of Fe$^{+2}$ moments by {\bf k} = (1,\,0,\,1) \cite{Xiao2009}. The appearance of strongest magnetic Bragg peak at the weakest nuclear Bragg peak (1\,0\,1) at $d=3.73$~\AA\ and the absence of incommensurate peaks further support the {\bf k} = (0,\,0,\,0) propagation vector.

\begin{figure}
\includegraphics[width=3in]{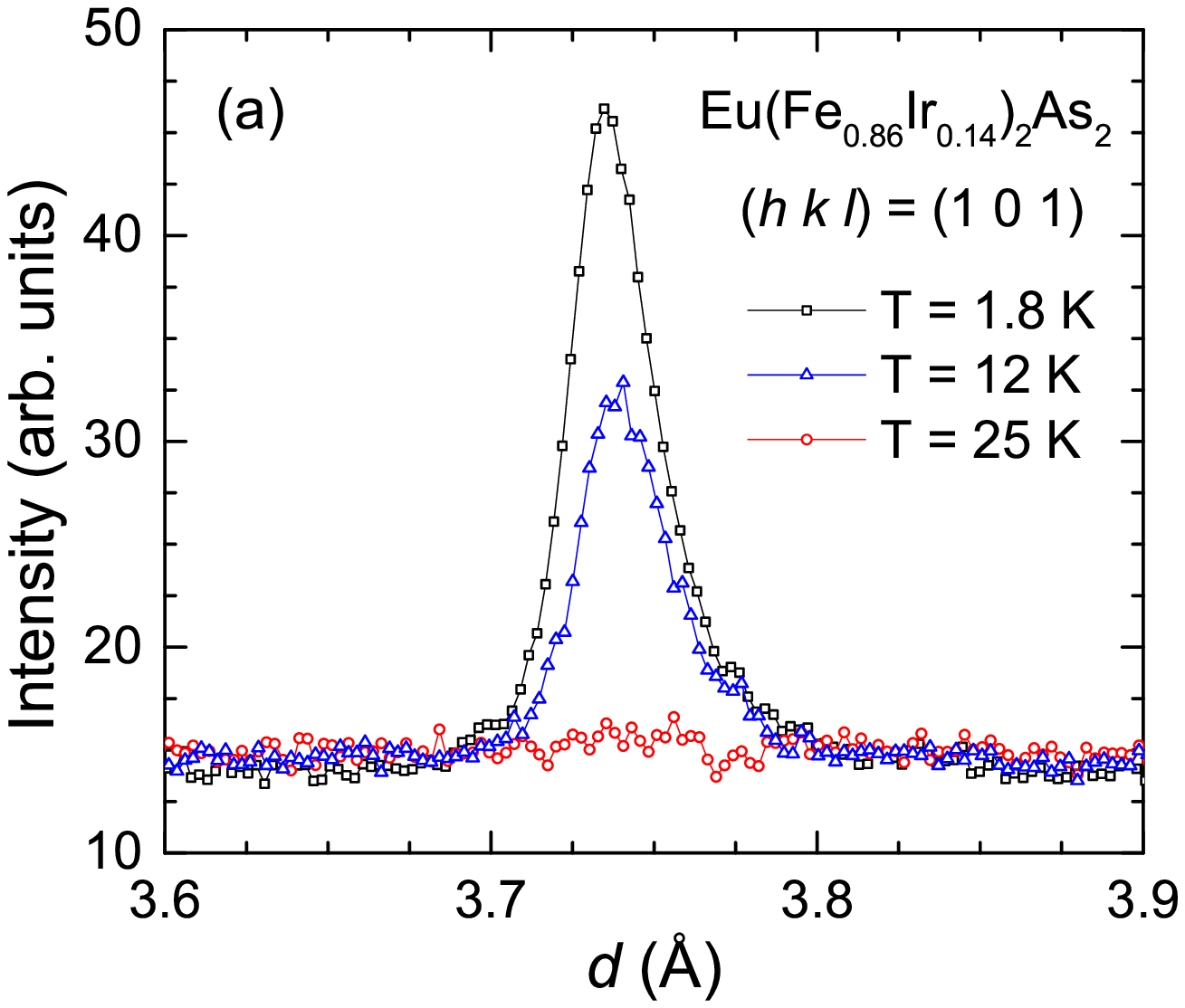}
\includegraphics[width=3in]{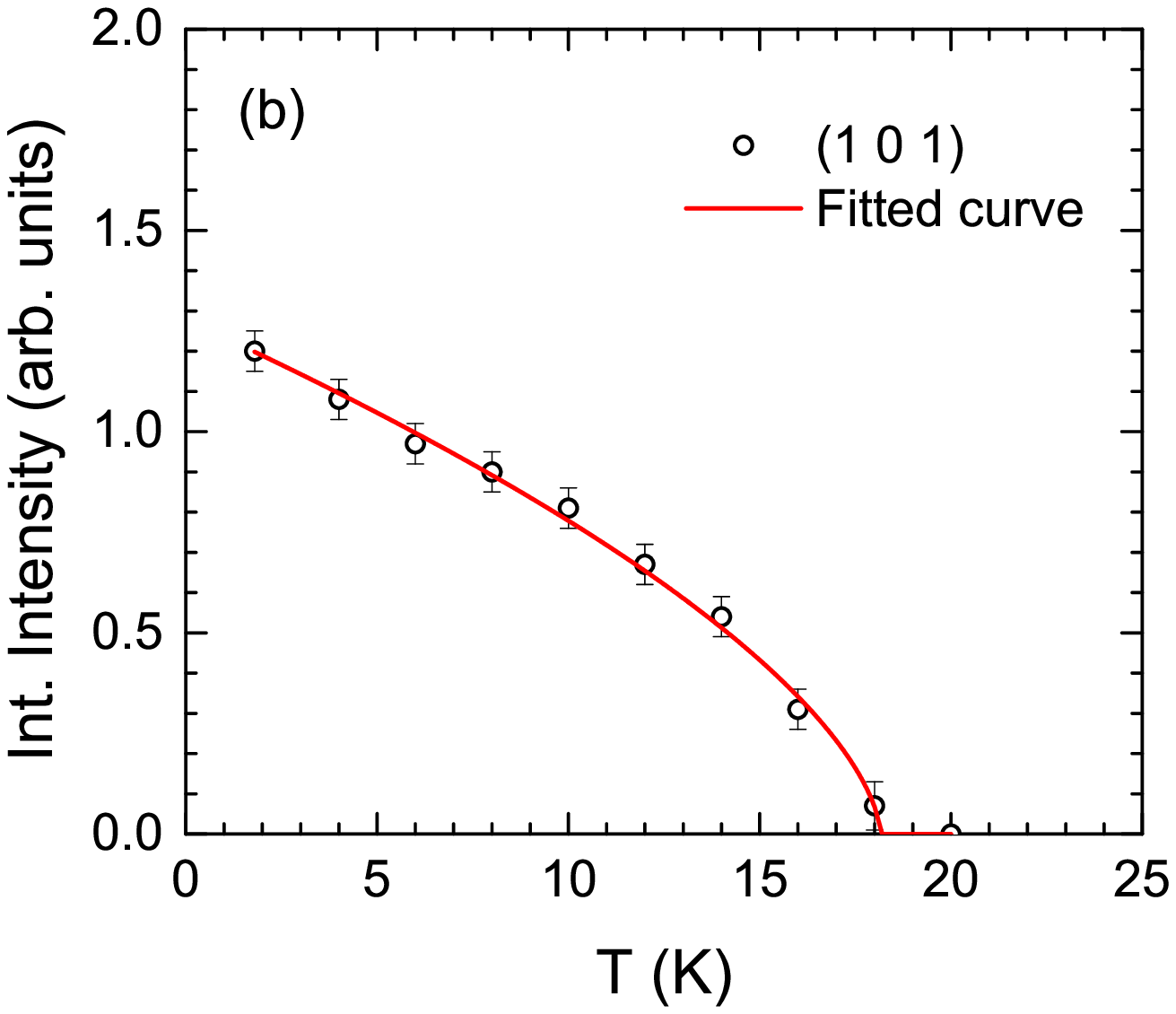}
\caption{\label{fig:ND2} (Color online) (a) Neutron diffraction Bragg peak (1\,0\,1) at temperatures above (25~K) and below (1.8~K, 12~K) magnetic ordering at $T_{\rm Eu} \approx 19$~K in Eu(Fe$_{0.86}$Ir$_{0.14}$)$_2$As$_2$. (b) Temperature $T$ dependence of the integrated intensity of the (1\,0\,1) magnetic Bragg peak. The solid curve represents the fit according to $I = I_0 (1 - T/T_{\rm Eu})^{2\beta}$ for $T_{\rm Eu} = 18.2(1)$~K and $\beta=0.31(2)$.}
\end{figure}

The intensity of observed magnetic peaks decreases with increasing temperature. The $T$ dependence of intensity is evident from Fig.~\ref{fig:ND2}(a) in which the intensity of (1\,0\,1) Bragg peak is compared for three temperatures 25~K ($> T_{\rm Eu}$) and 1.8~K and 12~K ($< T_{\rm Eu}$). The temperature evolution of integrated intensity of (1\,0\,1) magnetic Bragg peak is shown in Fig.~\ref{fig:ND2}(b). The $T$ dependent intensity of the magnetic peak represents the order parameter of the magnetic transition, which is well fitted by a power law $I \propto (1 - T/T_{\rm Eu})^{2\beta}$, yielding the transition temperature $T_{\rm Eu}$ = 18.2(1) K and the exponent $\beta = 0.31(2)$. $T_{\rm Eu}$ determined here is in good agreement with the results from $\mu$SR, resistivity and magnetization measurements. The $\beta$ obtained so is comparable to $\beta = 0.35(2)$ for Eu(Fe$_{0.82}$Co$_{0.18}$)$_2$As$_2$ \cite{Jin2013} and $\beta = 0.36(4)$ for EuFe$_{2}$(As$_{0.81}$P$_{0.19}$)$_{2}$ \cite{Nandi2014b} and to that of the three-dimensional Heisenberg system ($\beta \approx 0.36$) \cite{Hohenemser1989,Kagawa2005}.

\begin{figure}
\includegraphics[width=1.8in]{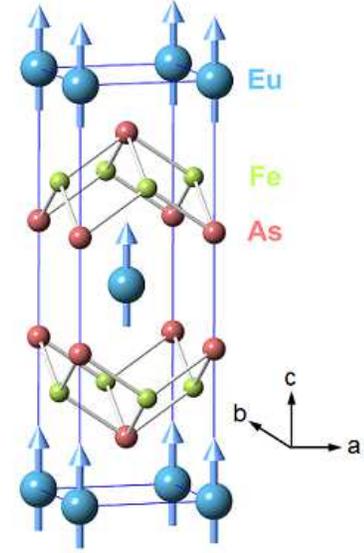}
\caption{\label{fig:Mag_struct} (Color online) The body centered tetragonal chemical and magnetic unit cell of Eu(Fe$_{0.86}$Ir$_{0.14}$)$_2$As$_2$ (space group $I4/mmm$). The arrows denote the ordered Eu$^{+2}$ magnetic moment directions. Ir atoms occupy the Fe site.}
\end{figure}

In order to determine the magnetic structure we employed a method whereby combinations of axial vectors localized on the 2a(Eu) site and also on the 4d(Fe) site and transforming as basis functions of the irreducible representations of the wave vector group [{\bf k} = (0,\,0,\,0)] are systematically tested. We carried out symmetry analysis using SARAh program \cite{Wills2000}. The symmetry analysis for the Eu moment ordering showed two non-zero irreducible representations (IRs), one is one-dimensional representation, labelled $\Gamma_3$  and the other is two-dimensional representation $\Gamma_9$ in the little group. Only $\Gamma_3$ and $\Gamma_9$ enter the decomposition of $\Gamma_{\rm mag\,Eu} = \Gamma_3 + \Gamma_9$.  $\Gamma_3$ corresponds to FM ordering of the Eu sites with moments along the $c$-axis, while $\Gamma_9$ corresponds to FM ordering of the Eu sites with moments along the $a$-axis or $b$-axis.

On the other hand, the symmetry analysis for the Fe moment ordering yielded that only four IRs enter the magnetic decomposition, two of which are one-dimensional representations (labelled $\Gamma_3$ and $\Gamma_6$) and remaining two are two-dimensional representations ($\Gamma_9$ and $\Gamma_{10}$) in the little group. Therefore $\Gamma_{\rm mag\,Fe} = \Gamma_3 + \Gamma_6 + \Gamma_9 + \Gamma_{10}$.  $\Gamma_3$ corresponds to FM ordering of the Fe sites with moments along the $c$-axis, whereas $\Gamma_6$ corresponds to AFM ordering along $c$-axis. $\Gamma_9$ shows FM along $a$-axis or FM along $b$-axis, while $\Gamma_{10}$ shows AFM along $a$-axis or AFM along $b$-axis.

In our analysis we tried to fit the magnetic structure using IRs $\Gamma_3$ and $\Gamma_9$. The analysis for $\Gamma_9$ did not fit the observed intensity near 5.94~\AA\ and 3.73~\AA\ well. The fit gave more intensity for 5.94~\AA\ peak than the observed one, while the calculated intensity of the 3.73~\AA\ peak was almost half of the observed intensity. This indicated that IR $\Gamma_9$ does not provide the correct representation for our data. Next we tried fitting our data based on IR $\Gamma_3$, which has one basis function. This basis function gives FM alignment of both Eu and Fe moment along the $c$-axis. First we allowed to vary both Eu and Fe moment and we obtained a reasonably good fit to the data with moment on Eu [$6.27(5)\,\mu_{\rm B}$] and Fe [$0.12(6)\,\mu_{\rm B}$] at 1.8~K with magnetic Bragg factor 12.6\%. In the second state of refinement we kept Fe moment fixed to zero and only varied Eu moment. A good fit of the 1.8~K ND data was obtained for Eu moment of $6.29(5)\,\mu_{\rm B}$ with magnetic Bragg factor 12.5\%. Though the magnetic Bragg factor is nearly the same with and without Fe moment fit, the large error bar on the Fe moment which is $\approx 50$\% of the observed Fe moment value of $0.12(6) \,\mu_{\rm B}$ may suggest that this is not a good solution and the Fe moment can be set to zero. The zero value of Fe moment was also observed in our M\"ossbauer study \cite{Paramanik2014}, which supports our analysis with fixing Fe moment to zero. The refinement of magnetic Bragg peaks at 1.8~K within IR $\Gamma_3$ is shown in Fig.~\ref{fig:ND1}(c). The magnetic structure of Eu(Fe$_{0.86}$Ir$_{0.14}$)$_2$As$_2$ determined this way consists of ferromagnetically coupled Eu moments aligned along the $c$-axis and is shown in Fig.~\ref{fig:Mag_struct}. The ordered moment $6.29(5)\,\mu_{\rm B}$  obtained from the refinement of 1.8~K neutron diffraction is somewhat smaller than the theoretically expected value of $7.0\,\mu_{\rm B}$ for Eu$^{+2}$ ($S = 7/2$) ions. An ordered moment of $6.6(2)\,\mu_{\rm B}$ was found from neutron diffraction study on EuFe$_{2}$(As$_{0.81}$P$_{0.19}$)$_{2}$ \cite{Nandi2014b}

\section{\label{Conclusion} Summary and Conclusions}

The interplay of magnetic order and superconductivity in Eu(Fe$_{0.86}$Ir$_{0.14}$)$_2$As$_2$ has been investigated by microscopic tools of $\mu$SR and neutron powder diffraction, and magnetic structure was determined. Our $\mu$SR data confirmed the long range magnetic ordering below $T_{\rm Eu} = 18.7(2)$~K and the magnetic ground state was found to be ferromagnetic in nature. The magnetic structure determination from neutron diffraction further confirmed the ferromagnetic structure showing that ferromagnetically coupled Eu moments are aligned along the $c$-axis with a magnetic propagation wavevector {\bf k} = (0,\,0,\,0) and ordered moment of $6.29(5)~\mu_{\rm B}$ at 1.8~K\@. Thus both $\mu$SR and neutron diffraction indicate ferromagnetic ordering in Eu(Fe$_{0.86}$Ir$_{0.14}$)$_2$As$_2$ which coexists with superconductivity as inferred from the magnetic susceptibility and resistivity measurements. As proposed earlier for the coexistence of ferromagnetic order and superconductivity in Eu(Fe$_{0.75}$Ru$_{0.25}$)$_2$As$_2$ \cite{Jiao2011} and EuFe$_{2}$(As$_{0.85}$P$_{0.15}$)$_{2}$ \cite{Nandi2014a}, the ferromagnetic superconductivity in the present compound may also be ascribed to the formation of spontaneous-vortex phase. Further investigations are desired to understand the mechanism of coexistence of magnetic order and superconductivity in these ferromagnetic superconductors.

It is seen that while substitution of Fe by Co, Ru and Ir leads to superconductivity, no superconductivity is observed in Ni-doped EuFe$_2$As$_2$. This contrasts the Ni doping in BaFe$_2$As$_2$ which also induces superconductivity and breaks the intuitive analogy between the doping behavior of EuFe$_2$As$_2$ and BaFe$_2$As$_2$. An obvious question is: Is it the ferromagnetic order of Eu$^{+2}$ moments which is responsible for the absence of superconductivity in Eu(Fe$_{1-x}$Ni$_{x}$)$_{2}$As$_{2}$? A comparative study of superconducting Co-, Ru- and Ir-doped EuFe$_2$As$_2$ and nonsuperconducting Ni-doped EuFe$_2$As$_2$ should prove informative in understanding the difference of the superconducting behaviors of these two groups of ferromagnetically ordered compounds and enrich our knowledge of the conditions to realize the coexistence of ferromagnetism and superconductivity.

Further, it appears that the substitution at Fe or As sites stabilizes the $ab$-plane ferromagnetic exchange interaction of undoped EuFe$_2$As$_2$ and the direction of Eu$^{+2}$ moments also changes from the $ab$-plane orientation to along the $c$-axis in doped EuFe$_2$As$_2$. The change in magnetic structure upon doping the Fe or As sites is most likely brought by the change in RKKY interaction which is mediated by the conduction electrons from the FeAs layers. If this is the case one would expect the nature of magnetic order to remain unchanged upon doping at Eu site or upon the application of pressure as the FeAs layers remain intact. It would be of interest to check this hypothesis by magnetic structure determination of Eu site doped EuFe$_2$As$_2$ and in the pressure induced superconducting state for which the coexistence of antiferromagnetic order and superconductivity is more intuitive.

\acknowledgments
We would like to thank Prof. A. M. Strydom for an interesting discussion. DTA and ADH acknowledge financial assistance from CMPC-STFC grant number CMPC-09108. AB thanks UJ and STFC for PDF funding. UBP and ZH acknowledge support from the Council of Scientific and Industrial Research (CSIR), New Delhi (Grant No.80(0080)/12/ EMR-II).

\end{document}